# Band flips and bound-state transitions in leaky-mode photonic lattices

Sun-Goo Lee and Robert Magnusson
Department of Electrical Engineering,
University of Texas at Arlington,
Arlington, Texas 76019, USA

**Leaky-mode photonic lattices exhibit intricate resonance effects originating in quasi-guided lateral Bloch modes. Key spectral properties are associated with phase-matched modes at the second (leaky) stop band. One band edge mode suffers radiation loss generating leaky-mode resonance whereas the other band edge mode becomes a bound state in the continuum (BIC). Here, we present analytical and numerical results on the formation and properties of the leaky stop band. We show that the frequency of the leaky-mode resonance band edge, and correspondingly the BIC edge, is determined by superposition of Bragg processes chiefly generated by the first two Fourier harmonics of the spatial modulation. We derive conditions for the band closure and band flip wherein the leaky edge and the bound-state edge transit across the band gap. Our work elucidates fundamental aspects of periodic photonic films and has high relevance to the burgeoning field of metamaterials.**

    Metamaterials constitute a new class of photonic platforms wherein principal performance metrics are controlled by the properties of a collection of subwavelength particles. Periodic and aperiodic metasurfaces and metagratings can be fashioned to provide complex functionality in extremely compact format even as single-layer films. Lossless dielectric media are particularly promising for high-efficiency applications [1]. Thus, there is great interest in exploring metamaterials as building blocks for high-performance photonic devices including metalenses [2], perfect reflectors [3], and meta-holograms [4]. Advances in theoretical modeling, numerical design methods, fabrication, and physical and spectral characterization are discussed in numerous recent review articles [5-8].

    Periodic subwavelength metastructures, including one-dimensional (1D) and two-dimensional (2D) metagratings in photonic-crystal slab geometry, are governed by principles that depend strongly on the scale of the operational wavelength $\lambda$ relative to the period $\Lambda$. In the deep subwavelength regime $\Lambda \ll \lambda$, classic effective-medium theory [9] becomes accurate and the material is effectively homogenized enabling facile phase control, anti-reflection properties, and polarization manipulation. In the subwavelength resonance regime with the period moderately smaller than the wavelength $\Lambda < \lambda$, effective medium theory fails on account of coupling of incident light to lateral leaky resonant modes. Such devices exhibit guided-mode resonance (GMR) effects caused by lateral Bloch modes in both 1D and 2D periodic lattices [10]. The resonance regime enables a great variety of novel device concepts including efficient wide-band reflectors, narrow bandpass filters, and polarizers [11].

    In this paper, we address fundamental properties of the photonic band structure of resonant leaky-mode metamaterials. The band structure admits a leaky edge and a non-leaky edge for each supported resonant Bloch mode if the lattice is symmetric. The non-leaky edge is associated with a bound state in the continuum (BIC), or embedded eigenvalue, currently of great scientific interest [12-21]. It is possible to control the width of the leaky band gap by lattice design. In particular, as a modal band closes, there results a quasi-degenerate state—this state is remarkable as it is possible to transit to it by parametric and material choice as shown in this paper. The transition to, and across, this point executes a band flip. The physical mechanisms inducing the band closure and the band flip have thus far not been explained. Hence, using semi-analytical and rigorous numerical methods, we characterize band flips and BICs relative to lattice harmonic content and device parameters. We treat a simple 1D photonic lattice supporting counterpropagating Bloch modes in a single polarization state. This canonical case brings forth all the principal properties and can readily be extended to 2D lattices. The band-flip concept provided here has relevance to general periodic photonic lattices and metamaterials that are of high interest in various branches of photonics as summarized above [1-8].

## Results
### Lattice structure and perspective

Figure 1 illustrates our simple model and the attendant schematic dispersion relations indicating the band flip. As noted in Fig. 1a, we analyse a single 1D periodic layer with thickness $d$ with binary dielectric-constant modulation enclosed by a substrate with dielectric constant $\varepsilon_s$ and a cover region of $\varepsilon_c$. The periodic layer acts as a waveguide as well as a phase-matching element because its average dielectric constant $\varepsilon_{avg} = \varepsilon_l + \rho(\varepsilon_h - \varepsilon_l)$ is larger than $\varepsilon_s$ and $\varepsilon_c$, where $\varepsilon_h$ and $\varepsilon_l$ represent the high and low dielectric constants, respectively, and where $\rho$ is the fill factor of the high dielectric constant part. The normally incident light is in the TE polarization state such that the electric field vector is along the $y$-direction. In this work, the average dielectric constant of the guiding layer is kept constant to highlight the effect of changes in index modulation clearly. We use a grating strength parameter $\Delta\varepsilon = \varepsilon_h - \varepsilon_l$ to represent the level of refractive-index modulation. In this 1D case, photonic band gaps open up for media with $\varepsilon_h$ and $\varepsilon_l$ when $0 < \rho < 1$ and $\Delta\varepsilon > 0$. As shown schematically in Fig. 1b, leaky-mode resonance or, equivalently, GMR reflectance peaks corresponding to transmittance (T) nulls appear at the lower edge of the second (leaky) stop band when the values of $\rho$ and $\Delta\varepsilon$ are small. The location of the resonance transits from the lower to the upper band edge as the values of $\rho$ or $\Delta\varepsilon$ increase with a corresponding transition of the BIC state as indicated. In general, if the lattice supports numerous leaky modes, each mode will undergo similar band flips as each mode possess a band gap [22]. In this paper, we limit our attention to the fundamental TE mode as this simplest case brings out the key properties of the band-flip effect.

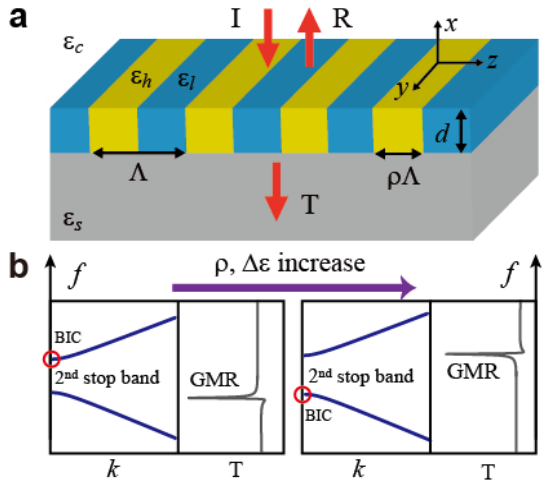

**Figure 1 | Band flip in a leaky-mode resonant photonic lattice. a,** Schematic of a resonant lattice with a normally-incident TE-polarized plane wave. **b,** Conceptual illustration of the band flip phenomenon. When the values of $\rho$ and $\Delta\varepsilon$ are small, GMR (BIC in a red circle) occurs at the lower (upper) side of the second stop band. The band flip refers to the transition of the GMR (BIC) location from lower (upper) to upper (lower) band edge as $\rho$ and $\Delta\varepsilon$ increase.

### Semi-analytical dispersion model

Dispersion relations pertinent to guided modes in periodic lattices can be obtained from the wave equation [23]

$$\nabla^2 E(\mathbf{r}) - \left(\frac{2\pi f}{c}\right)^2 \varepsilon(\mathbf{r}) E(\mathbf{r}) = 0, \qquad (1)$$

where $c$ is the speed of light in free space and $f$ denotes frequency. To solve the equation, the periodic dielectric function $\varepsilon(\mathbf{r})$ is expanded in a Fourier series and the electric field is expanded as plane waves as $E(\mathbf{r}) = \sum_{n=-\infty}^{\infty} E_n e^{i(k+n\mathrm{K})r}$ where $\mathrm{K} = 2\pi/\Lambda$ is the magnitude of the grating vector and $k$ is wavenumber [24]. For the 1D symmetric lattice, the dielectric function can be expanded in an even cosine function series $\varepsilon(z) = \sum_{n=0}^{\infty} \varepsilon_n \cos(n\text{K}z)$ with Fourier coefficients given by $\varepsilon_0 = \varepsilon_{avg}$ and $\varepsilon_{n\geq 1} = (2\Delta\varepsilon/n\pi)\sin(n\pi\rho)$.

Central to our results is the dispersion relation relating frequency of light to its wavenumber thus determining the propagation properties of light in media [25-30]. For clear insight, we use a semi-analytical dispersion model of the second stop band to study the band transitions; this model is then verified by rigorous finite-difference time-domain (FDTD) computations. Kazarinov and Henry (KH) employed coupled-mode theory to study the leaky stop band. For lattices with symmetric profiles, the KH model implies one leaky band edge whereas the other one is non-leaky [28]. In an alternative view, the non-leaky band edge can be understood as a BIC or symmetry-protected state. These states have been intensively studied because they can achieve localized states with infinite lifetimes implying, in principle, that there is no radiation decay [12-21].

The KH model solves the wave equation semi-analytically by retaining only the zeroth, first, and second Fourier harmonics. The spatial electric field distribution is approximated as $E(x,z) = [A\exp(i\text{K}z) + B\exp(-i\text{K}z)]\varphi(x) + E_{rad}$, where $\varphi(x)$ characterizes the mode profile of the unmodulated waveguide and $E_{rad}$ represents the radiating diffracted wave. Near the second stop band, the dispersion relation can be written as

$$\Omega(k) = \Omega_0 \pm \sqrt{k^2 + (h_2 + ih_1)^2}/(\text{K}h_3) - ih_1/(\text{K}h_3) \quad (2)$$

where $\Omega = \Omega_{Re} + i\Omega_{Im} = 2\pi f/\text{K}c$ represents the normalized frequency and $\Omega_0$ is the Bragg frequency under vanishing index modulation (~ homogeneous waveguide) and $k$ is the Bloch wave vector in the periodically modulated waveguide [28,29]. The coefficient $h_1$ represents the coupling between a guided wave and a radiated wave, $h_2$ denotes coupling between two counter-propagating lateral waveguide modes and $h_3$ is a coefficient related to the group velocity in the unmodulated waveguide with dielectric constant of $\varepsilon_{avg}$. For TE modes, the three coefficients are given by [29,30]

$$\begin{bmatrix} h_1 = i\dfrac{\text{K}^3\Omega^4\varepsilon_1^2}{8}\int_{-d}^{0}\int_{-d}^{0} G(x,x')\varphi(x')\varphi^*(x)\,dx'dx \\ h_2 = \dfrac{\text{K}\Omega^2\varepsilon_2}{4}\int_{-d}^{0}\varphi(x)\varphi^*(x)\,dx \\ h_3 = \Omega\int_{-\infty}^{\infty}\varepsilon_0(x)\varphi(x)\varphi^*(x)\,dx \end{bmatrix} \quad (3)$$

where $G(x,x')$ denotes the Green's function for the diffracted field; see Supplementary Materials. As indicated in equation (2), the KH model yields two different frequencies for a given $k$. The leaky stop band with two band edges $\Omega^+ = \Omega_0 + h_2/(\text{K}h_3)$ and $\Omega^- = \Omega_0 - (h_2 + i2h_1)/(\text{K}h_3)$ opens at $k = 0$. At the band edge with frequency $\Omega^+$ obtained when the electric field distribution is an antisymmetric (sine) function ($A = -B$), there is no radiation loss because $\Omega^+$ is purely real. At the $\Omega^-$ band edge obtained when the field distribution is a symmetric (cosine) function ($A = B$), the radiative loss is maximal with $\text{Im}(\Omega^-) = -2\text{Re}(h_1)/(\text{K}h_3)$. Hence, the band edge modes with the frequencies $\Omega^-$ and $\Omega^+$ are associated with GMR and BIC, respectively.

**Bragg-reflection superposition model of the leaky stop band**

We now show that the frequency location of the leaky-mode resonance band edge, or the BIC edge, is determined by superposition of Bragg processes denoted by $\text{BR}_{q,n}$ where $q$ indicates the Bragg order and $n$ denotes the Fourier harmonic of the dielectric constant modulation. As an approximation, we keep only the strongest Bragg processes which are $\text{BR}_{2,1}$ operating as a second-order Bragg reflection off the first Fourier harmonic and $\text{BR}_{1,2}$ defining a first-order Bragg reflection by the second harmonic. The Bragg reflection superposition model proposed here is based on the fact that the size of the second stop band is given by $|\text{Re}(\Omega^+) - \text{Re}(\Omega^-)| = 2|h_2 - \text{Im}(h_1)|/(\text{K}h_3)$ from equation (2) with coupling coefficients $h_1$ and $h_2$ related to the first and the second Fourier coefficients, respectively, as seen in equation (3). The band gap will disappear when $h_2 = \text{Im}(h_1)$ because the two Bragg reflections $BR_{2,1}$ and $BR_{1,2}$ are then balanced destructively.

To accurately evaluate stop band formation by these Bragg processes, we calculate the band structures of pertinent 1D lattices by FDTD simulations [30,31]. In Fig. 2a, for a representative set of lattice parameters provided in the figure caption, the stop band denoted $\Delta\Omega_1$ is shown for a lattice having only the fundamental harmonic such that $\varepsilon(z) = \varepsilon_0 + \varepsilon_1 \cos(Kz)$. Dispersion curves (blue lines) obtained from the full non-approximated lattice are also plotted for comparison. Clearly, the FDTD results with the fundamental harmonic only are quite different from those with the full lattice. Figure 2b shows stop band $\Delta\Omega_2$ formed by a first-order scattering process off the second harmonic. The full-lattice band structure is close to the approximate structure denoting the importance of this partial scattering process. Figure 2c shows that the third order harmonic $\varepsilon_3 \cos(3Kz)$ cannot contribute to the second stop band by itself. Figure 2d illustrates that the band $\Delta\Omega_{12}$ simulated with the first and second harmonics simultaneously agrees well with the band $\Delta\Omega$ simulated with the full non-approximated lattice. Moreover, there is excellent agreement with the dispersion curves calculated with the KH model. Hence, it is reasonable to conclude that the Bragg-reflection superposition model proposed here is valid to describe the second stop band of weakly to moderately modulated photonic lattices.

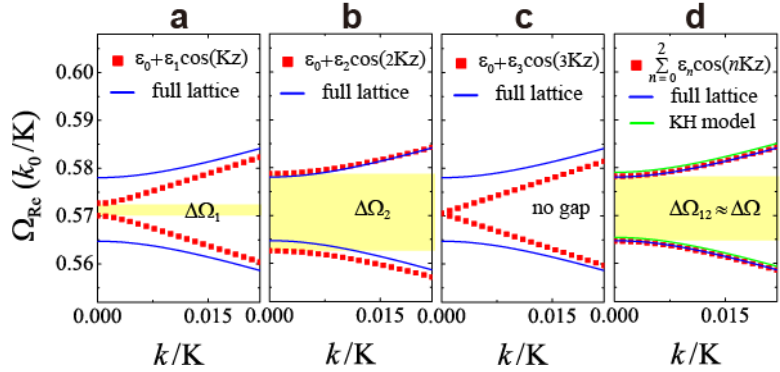

**Figure 2 | Computed stop bands for a 1D leaky-mode lattice relative to Fourier harmonic content.** The dielectric functions vary for these examples. **a,** $\varepsilon = \varepsilon_0 + \varepsilon_1 \cos(Kz)$, **b,** $\varepsilon = \varepsilon_0 + \varepsilon_2 \cos(2Kz)$, and **c,** $\varepsilon = \varepsilon_0 + \varepsilon_3 \cos(3Kz)$. In **d** $\varepsilon = \varepsilon_0 + \varepsilon_1 \cos(Kz) + \varepsilon_2 \cos(2Kz)$ is used. Parameters for the FDTD simulations and KH model are $d = 0.50\Lambda$, $\rho = 0.35$, $\varepsilon_c = 1.00$, $\varepsilon_s = 2.25$, $\Delta\varepsilon = 1.00$, and $\varepsilon_{avg} = 4.00$.

Conditions for the constructive or destructive interaction between the $BR_{2,1}$ and $BR_{1,2}$ scattering processes can be inferred from the coupling coefficients $h_1$ and $h_2$. For the lattice shown in Fig. 1a, $h_2$ is positive (negative) when the fill factor $\rho$ is smaller (greater) than 0.5. But $\text{Im}(h_1)$ is always positive irrespective of $\rho$. Since the size of the second band gap is proportional to $|h_2 - \text{Im}(h_1)|$, when $\rho > 0.5$ with $h_2 < 0$, the size of the gap results from the addition of the two coupling coefficients. When $\rho < 0.5$, on the other hand, the gap size is determined by the subtraction of the two coefficients and thus the band gap can reach a zero value.

**Symmetry properties of the band-edge modes**

The physics of band formation and band flips can be further understood from the electric field distributions of the band edge modes computed with FDTD simulations. At the edges of $\Delta\Omega_1$ with $\varepsilon(z) = \varepsilon_0 + \varepsilon_1 \cos(Kz)$ as shown in Figs. 3a,c the lower (upper) band edge modes have antisymmetric (symmetric) field ($E_y$) distributions irrespective of the value of $\rho$. At the edges of band $\Delta\Omega_2$ with $\varepsilon(z) = \varepsilon_0 + \varepsilon_2 \cos(2Kz)$, on the other hand, electric field distributions are dependent on the value of the fill factor because the second Fourier harmonic $\varepsilon_2 = (\Delta\varepsilon/\pi)\sin(2\pi\rho)$ changes its sign once from + to

– when $\rho = 0.5$. Figure 3 shows that the $\Delta\Omega_1$ and $\Delta\Omega_2$ bands have field distributions with matched symmetry at both edges for $\rho > 0.5$. This explains why GMR always appears at the upper edge when $\rho > 0.5$. Here, the field patterns pertaining to the upper edge are symmetric (Fig. 3c,d) allowing constructive interaction, radiation, and resonance response. Moreover, this symmetry state is stable as $\Delta\varepsilon$ increases such that the resonance remains at the upper edge irrespective of modulation level. In contrast, the lower edge antisymmetric field structure (Fig. 3a,b) cannot be excited at normal incidence; hence, there can be no resonance generation and the state is a symmetry-protected BIC. We can interpret processes $BR_{2,1}$ and $BR_{1,2}$ forming the two bands $\Delta\Omega_1$ and $\Delta\Omega_2$ as interacting destructively (constructively) when $\rho < 0.5$ ($\rho > 0.5$) because the field distributions are opposite (same). As destructive interaction between the processes closes the gap and induces band flip, $\rho < 0.5$ is the important case. In this case, it is possible to alter the field-symmetry structure via the $BR_{q,n}$ processes by varying the level of modulation.

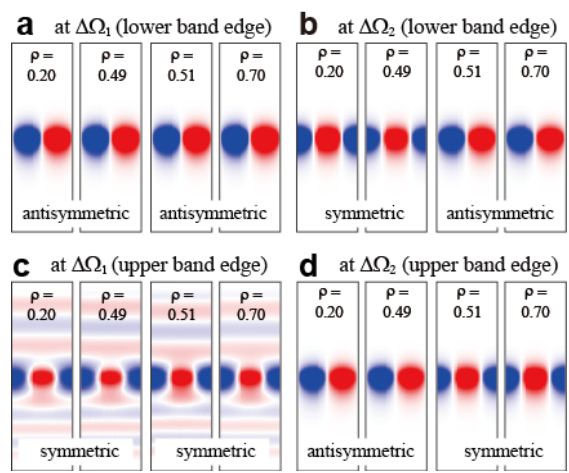

**Figure 3 | Spatial electric field distributions at band edges.** Simulated electric field ($E_y$) distributions at the lower band edges of **a**, $\Delta\Omega_1$ and **b**, $\Delta\Omega_2$ and at the upper band edges of **c**, $\Delta\Omega_1$ and **d**, $\Delta\Omega_2$. At $\Delta\Omega_1$ that is formed via process $BR_{2,1}$ symmetries of electric field distributions are irrespective of $\rho$. At $\Delta\Omega_2$ formed by process $BR_{1,2}$, on the other hand, the field distributions are reversed once due to the sign change of the second Fourier coefficient. Lattice parameters are the same as in Fig. 2 except that $\rho$ varies.

**Band evolution relative to modulation strength**

Whereas the field distributions for $\rho > 0.5$ are symmetry matched, the distributions for $\rho < 0.5$ are mismatched as seen in Fig. 3. A physical resonance at normal incidence demands local field symmetry thus locating at the corresponding edge with the BIC forced to appear at the other edge. Here, the electric field distributions at the leaky stop band $\Delta\Omega$ are determined by the competition between Bragg processes $BR_{2,1}$ and $BR_{1,2}$. When both $\rho$ and $\Delta\varepsilon$ are small, the non-leaky antisymmetric (leaky symmetric) mode will locate at the upper (lower) band edge because the first order reflection $BR_{1,2}$ suppresses the second order reflection $BR_{2,1}$. But when $\rho$ increases and approaches 0.5, there is a chance for $BR_{2,1}$ to overwhelm $BR_{1,2}$ because the strength of $BR_{1,2}$ gets weaker and becomes zero as $\varepsilon_2$ approaches zero. For a given value of $\rho$ ($< 0.5$), as $\Delta\varepsilon$ increases from zero there should exist a critical value of index modulation $\Delta\varepsilon_{BF}$ where the band gap closes with $h_2 = \text{Im}(h_1)$. Before (After) the band gap closure, GMRs should appear at lower (upper) band edges. As the value of $\rho$ gets closer to 0.5, a smaller value of index modulation will be required for $BR_{2,1}$ and $BR_{1,2}$ to balance each other because the coupling coefficients $h_1$ and $h_2$ are proportional to $[\Delta\varepsilon \times \sin(\pi\rho)]^2$ and $\Delta\varepsilon \times \sin(2\pi\rho)$, respectively.

To complement the insights gained from the KH model and the full FDTD computations, we consider the simpler band structure of a half-wave layer stack of infinite lateral extent. As the structure is fully transmissive there is no reflection and the band is closed. Light with wavelength $\lambda$ which satisfies the condition $(\lambda/2) = t_h = t_l$, where $t_h = \rho\Lambda\sqrt{\varepsilon_h}$ and $t_l = (1-\rho)\Lambda\sqrt{\varepsilon_l}$ denote optical thickness of high and low index layers, respectively, transmits through each layer without reflection due to consecutive Fabry-Perot resonances. From this condition, we find the relation $\rho = \sqrt{\varepsilon_l}/(\sqrt{\varepsilon_h} + \sqrt{\varepsilon_l})$ when the band gap vanishes. For the leaky-mode resonant lattice with finite thickness under study here, we can by analogy infer the condition for the band gap closure $L_h = L_l$ or $\rho = N_l/(N_h + N_l)$, where $N_h$ and $N_l$ represents the effective refractive index of the guided mode in a uniform waveguide with dielectric constant $\varepsilon_h$ and $\varepsilon_l$, respectively, and $L_h = \rho\Lambda N_h$ and $L_l = (1-\rho)\Lambda N_l$ are attendant effective optical path lengths. Effective indices $N_h(\neq \sqrt{\varepsilon_h})$ and $N_l(\neq \sqrt{\varepsilon_l})$ are functions of frequency due to waveguide dispersion. Before ($L_h < L_l$) and after ($L_h > L_l$) the band closure, GMRs appear at the lower and upper band edge, respectively.

**Table 1 | Simulated lattice modulation $\Delta\varepsilon_{BF}$ at band flip.** As $\rho$ decreases from 0.5, the modulation strength $\Delta\varepsilon_{BF}$ increases. The FDTD-simulated $\Delta\varepsilon_{BF}$ are used to calculate $\rho_{BF} = N_l/(N_h + N_l)$. Here, $N_h$ and $N_l$ are calculated from waveguide theory by employing the relations $\varepsilon_h = \varepsilon_{avg} + (1-\rho)\Delta\varepsilon_{BF}$ and $\varepsilon_l = \varepsilon_{avg} - \rho\Delta\varepsilon_{BF}$, respectively.

| $\rho$ | 0.49 | 0.48 | 0.47 | 0.46 | 0.45 |
|---|---|---|---|---|---|
| $\Delta\varepsilon_{BF}$ | 0.31 | 0.62 | 0.93 | 1.24 | 1.55 |
| $\rho_{BF}$ | 0.490 | 0.480 | 0.471 | 0.461 | 0.452 |

In order to verify the dependence of $\Delta\varepsilon_{BF}$ on $\rho$, we investigate $\Delta\varepsilon_{BF}$ through FDTD simulations with results for five different fill factors given in Table 1. It is seen that $\Delta\varepsilon_{BF}$ increases from 0.31 to 1.55 when $\rho$ decreases from 0.49 to 0.45. In the simulations, we use the full non-approximated lattice and $\Delta\varepsilon$ is increased in the step of 0.01. We next check the validity of the relation $\rho_{BF} = N_l/(N_h + N_l)$ inferred from the half-wave stack. By employing the simulated values of $\Delta\varepsilon_{BF}$ for the given $\rho$ shown in Table. 1, effective indices $N_h$ and $N_l$ are first calculated via classic dielectric waveguide theory with $\rho_{BF}$ found subsequently. Table 1 reveals that the calculated $\rho_{BF}$ via the half-wave model agrees well with the input values of $\rho$. Hence, we conclude that the relation for the band gap closure $\rho_{BF} = N_l/(N_h + N_l)$ is a good approximation.

Figure 4a shows the evolution of the leaky-mode stop band under variation of $\Delta\varepsilon$ for $\rho = 0.48$. As $\Delta\varepsilon$ increases from zero, the band gap opens and its size increases. However, the gap decreases and becomes zero as $\Delta\varepsilon$ is further increased. On additional increase in $\Delta\varepsilon$ the band gap reopens and its size grows again. These dynamics are associated with band flip as seen by investigating the complex part of the frequency $\Omega_{Im}$ shown in Fig. 4b. Since the time dependence of the electric field as $\exp(-i\Omega t)$ is employed in our FDTD simulations, the time dependent modal radiation loss is represented by $\Omega_{Im}$ with negative values. Figure 4b shows that the non-leaky (leaky) band edge mode associated with BIC (GMR) transits from the upper (lower) to the lower (upper) side of the stop band as $\Delta\varepsilon$ increases.

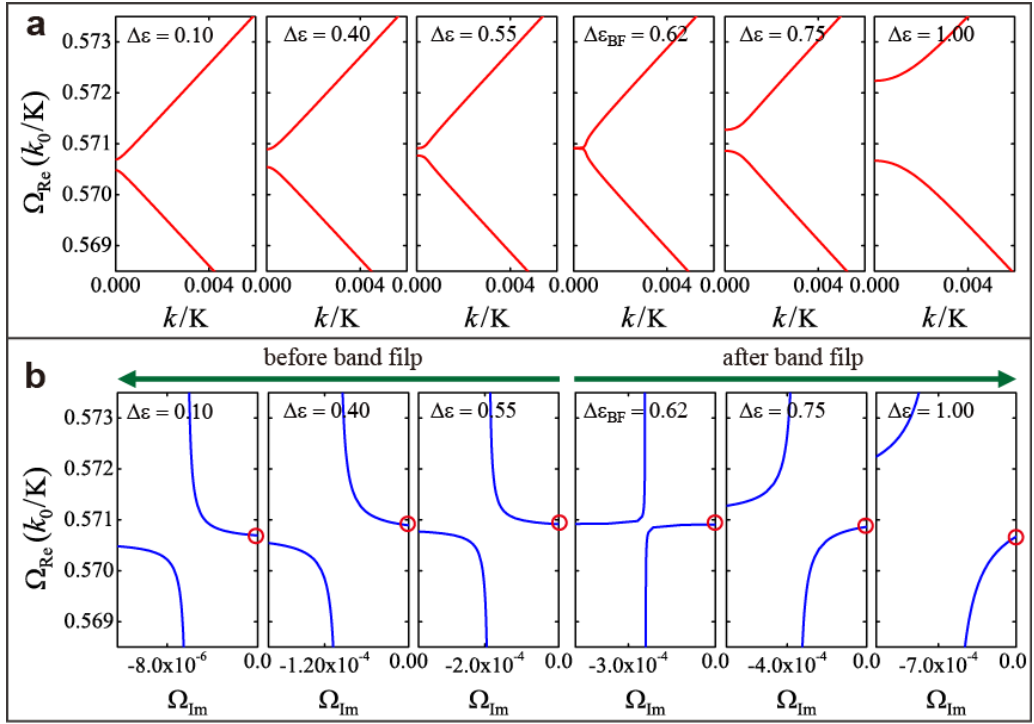

**Figure 4 | Evolution of photonic bands under variation of index modulation. a,** Simulated dispersion relations near the second stop band for six different values of $\Delta\varepsilon$. The band gap closes when $\Delta\varepsilon_{BF} = 0.62$. **b,** Relations between the imaginary part $\Omega_{Im}$ and real part $\Omega_{Re}$ of the frequency. Red circles indicate the non-leaky BIC band edges. In the FDTD simulations, we used structural parameters $d = 0.50\Lambda$, $\rho = 0.48$, $\varepsilon_c = 1.00$, $\varepsilon_s = 2.25$, and $\varepsilon_{avg} = 4.00$.

**Leaky band flattening**

Out-of-plane radiation at the leaky edge with a bound state at the opposing edge are primary aspects of the photonic lattices under study herein. Band flips and bound-state transitions have been formulated above. An additional interesting finding from these studies is that there exist a finite range of Bloch wave vectors $\Delta k$ where $\partial\Omega_{Re}/\partial k$ becomes zero as seen in Fig. 4a with $\Delta\varepsilon_{BF} = 0.62$ and shown enlarged in Fig. 5a. When the band gap closes with $h_2 = \text{Im}(h_1)$, the dispersion relation in equation (2) can be rewritten as

$$\Omega_{BF}(k) = \Omega_0 + \text{Im}(h_1)/(Kh_3) \pm \sqrt{k^2 - \text{Re}(h_1)^2}/(Kh_3) - i\,\text{Re}(h_1)/(Kh_3). \tag{4}$$

Equation (4) shows that the existence of the out-of-plane radiative loss flattens the dispersion curves because $\text{Re}[\Omega_{BF}^+(k)] = \text{Re}[\Omega_{BF}^-(k)] = \Omega_0 + \text{Im}(h_1)/(Kh_3)$ when $k^2 < \text{Re}(h_1)^2$. Comparing to the half-wave stack model, the band gap closes when $\rho = \sqrt{\varepsilon_l}/(\sqrt{\varepsilon_h} + \sqrt{\varepsilon_l})$ as shown above. However, Fig. 5b shows that the dispersion curves cross as straight lines and $\partial\Omega_{Re}/\partial k \neq 0$ at $k=0$ when the band closes. Recently, the linear dispersion associated with the stack model, called Dirac cone dispersion, has attracted much attention because a 2D photonic lattice with a Dirac cone at $k=0$ can act as a zero-refractive-index metamaterial with unusual light propagation properties [33,34]. It is the out-of-plane radiative loss associated with the leaky-mode lattice that flattens the band and prevents formation of a Dirac-cone-type band structure as seen in Fig. 5.

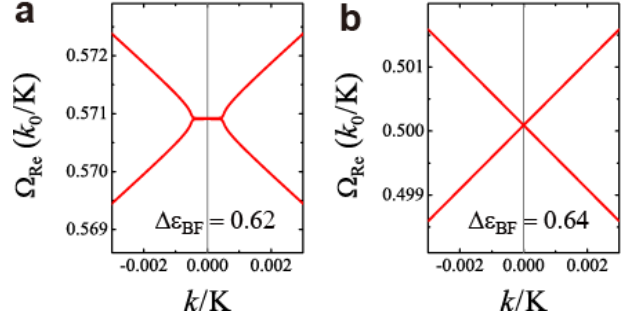

**Figure 5 | Dispersion relations at band gap closure. a,** Magnified dispersion relations for a leaky-mode lattice with $\Delta\varepsilon_{BF} = 0.62$ shown in Fig. 4a. The out-of-plane radiation loss flattens the dispersion curves in a finite range of Bloch wave vectors. **b,** Simulated dispersion relations for a non-leaky half-wave photonic lattice with $\rho = 0.48$ and $\varepsilon_{avg} = 4.00$. When $\Delta\varepsilon_{BF} = 0.64$, the band gap closes and two straight-line dispersion curves cross. The band closes when the relation $\rho = \sqrt{\varepsilon_l}/(\sqrt{\varepsilon_h} + \sqrt{\varepsilon_l})$ is satisfied.

**Discussion**

Our analysis of the band structure of leaky-mode photonic lattices shows that the band gap is primarily controlled by first-order Bragg diffraction by the second Fourier harmonic lattice component. However, near fill-factor of 0.5, second-order Bragg diffraction by the fundamental Fourier harmonic becomes competitive with the primary process. It is the destructive interference of these major processes that closes the gap and induces a band flip whereby the leaky edge and the bound-state edge transit across the band gap. Therefore, the band does not close at fill factor being identically 0.5 as often assumed. As the grating modulation strength increases, the transition point is increasingly pulled away from this value.

For this work, we employ a semi-analytic diffraction model due to Kazarinov and Henry [28]. Even though it requires determination of Green's functions, it provides direct insight into the operative processes by identification of coupling coefficients and their explicit parametric connections to the lattice parameters. Thus, the physical explanations can be argued straightforwardly and convincingly. We verify all approximate computations with simulations of the full lattice using rigorous FDTD models. Additionally, we make interesting connections with a half-wave stack lattice that is fully transmissive at band closure; this further informs the fundamental band properties of the leaky-mode lattice.

This research is limited to the simplest possible 1D lattice under weak or moderate levels of spatial modulation. Nevertheless, the essential band properties including resonance-edge and BIC-edge band flip, edge-mode electric-field structure, symmetry properties, and leaky-band flattening are brought out. Other lattices with arbitrary unit-cell structure and harmonic content can be treated analogously. The extension of this work to 2D lateral modulation as in photonic-crystal slabs is also feasible with many new insights expected. Thereby fundamental aspects of the band properties of 2D periodic photonic films and metamaterial structures can be understood.

**Methods**

The rigorous results provided in this article are computed using FDTD simulations. We use the semi-analytical KH model to obtain approximate solutions and to gain physical insight into the operative physics of the leaky-mode lattice under study. Further details are presented under Supplementary Materials.

**Acknowledgements**

This work is supported by the Texas Instruments Distinguished University Chair in Nanoelectronics endowment.

**Author contributions**

R.M. conceived the original idea. S.G.L conducted modelling and performed all numerical computations. The authors prepared the manuscript jointly.

## Supplementary materials: Methods

**Semi-analytical calculation of dispersion relations.**
To obtain the dispersion curves in Fig. 2d from the KH model, the grating layer shown in Fig. 1a is treated as an approximate homogeneous waveguide with a dielectric constant $\varepsilon_{avg}=4$. Three coupling coefficients $h_1$, $h_2$, and $h_3$ in equation (3) are numerically calculated by employing the mode profile $\varphi(x)$ and Green's function $G(x,x')$, summarized in the table below, for the three layer system composed of a cover (1), waveguide (2), and substrate (3) [30].

| Layer | $\varphi(x)$ | $G(x,x')$ |
|---|---|---|
| 1 | $Ne^{-\gamma_1 x}$ | $\dfrac{t_{21}(e^{-ik_2 x'}+r_{23}e^{2ik_2 d}e^{ik_2 x'})}{2ik_2(1-r_{21}r_{23}e^{2ik_2 d})}e^{ik_1 x}$ |
| 2 | $N(Ae^{ik_2 x}+Be^{-ik_2 x})$ | $\dfrac{(e^{-ik_2 x'}+r_{23}e^{2ik_2 d}e^{ik_2 x'})}{2ik_2(1-r_{21}r_{23}e^{2ik_2 d})}e^{ik_2 x}+\dfrac{r_{21}(e^{-ik_2 x'}+r_{23}e^{2ik_2 d}e^{ik_2 x'})}{2ik_2(1-r_{21}r_{23}e^{2ik_2 d})}e^{-ik_2 x}$, $(0 \geq x > x')$ <br><br> $\dfrac{r_{23}e^{2ik_2 d}(r_{21}e^{-ik_2 x'}+e^{ik_2 x'})}{2ik_2(1-r_{21}r_{23}e^{2ik_2 d})}e^{ik_2 x}+\dfrac{(r_{21}e^{-ik_2 x'}+e^{ik_2 x'})}{2ik_2(1-r_{21}r_{23}e^{2ik_2 d})}e^{-ik_2 x}$, $(x' \geq x > -d)$ |
| 3 | $NCe^{\gamma_3 x}$ | $\dfrac{t_{23}e^{i(k_2-k_3)d}(r_{21}e^{-ik_2 x'}+e^{ik_2 x'})}{2ik_2(1-r_{21}r_{23}e^{2ik_2 d})}e^{-ik_3 x}$ |

Here $k_i=-i\gamma_i=\sqrt{\varepsilon_i k_0^2-\beta_i^2}$ represents the wave vector component along the $x$-direction in each layer, where $\beta_i$ is the wave vector component along the lateral $z$-direction. $N^2=2k_2^2/[(1/\gamma_1+1/\gamma_3+d)+(k_2^2+\gamma_1^2)]$ for the normalization condition $\int_{-\infty}^{\infty}\varphi(x)\varphi^*(x)dx=1$. Coefficients $A=(k_2+i\gamma_1)/2k_2$, $B=(k_2-i\gamma_1)/2k_2$, and $C=(k_2+i\gamma_1)/(k_2-i\gamma_3)e^{(\gamma_3-ik_2)d}$ are related to the eigenvalue equation $k_2 d=m\pi+\tan^{-1}(\gamma_1/k_2)+\tan^{-1}(\gamma_3/k_2)$ for the modal dispersion of the $m$'th order mode. Coefficients $t_{ij}=2k_i/(k_i+k_j)$ and $r_{ij}=(k_i-k_j)/(k_i+k_j)$ are the transmission and reflection coefficients from the layer $i$ to layer $j$. As shown in equation (3), the coupling coefficients $h_1$, $h_2$, and $h_3$ are frequency dependent. However, for convenience, in this study the coefficients are calculated at a frequency of $\Omega_0=0.571$, the center of the second stop band estimated from the dispersion properties of the equivalent homogeneous waveguide, and assumed to be frequency independent. The assumption for the frequency independent coupling coefficients are appropriate at near $k=0$ for small index modulation because the frequency under consideration varies across a limited range [29].

**Finite-difference time-domain simulations.**
The simulations are carried out using a freely available software package MEEP [32]. Since the photonic structure studied here possesses a continuous symmetry in the $y$-direction, simulations are performed in a two-dimensional $xz$-plane. Computational cell of size $\Lambda \times 8\Lambda$, where $\Lambda$ is the grating period, is employed to obtain complex frequency $\Omega(k)$ and spatial electric field distributions. For reliable and stable simulations, spatial resolution and discrete time step are set to $\Delta x=\Delta z=\Lambda/50$ and $\Delta t=0.5\times \Delta x/c$, where $c$ is the speed of light in free space. Periodic and perfectly matched layer boundary conditions are used in the $x$- and $z$-directions, respectively.